# LIQUID CRYSTAL INDUCED ELASTO-CAPILLARY SUPPRESSION OF CRACK FORMATION IN THIN COLLOIDAL FILMS

*Udita Uday Ghosh[1], Jayabrata Dhar[2], Suman Chakraborty[2]\* and Sunando DasGupta[1]\**

1. *Microscale Transport Processes Laboratory, Department of Chemical Engineering, Indian Institute of Technology Kharagpur, India.*

2. *Microfluidics Laboratory, Department of Mechanical Engineering, Indian Institute of Technology Kharagpur, India.*

\*Corresponding author

Email:   sunando@che.iitkgp.ernet.in; suman@mech.iitkgp.ernet.in

**ABSTRACT**

Drying of colloidal droplets on solid, rigid substrates is associated with a capillary pressure developing within the droplet. In due course of time, the capillary pressure builds up due to droplet evaporation resulting in the formation of a colloidal thin film that is prone to crack formation. In this study, we show that introducing a minimal amount of nematic liquid crystal (NLC) can completely suppress the crack formation. The mechanism behind the curbing of the crack formation may be attributed to the capillary stress-absorbing cushion provided by the elastic arrangements of the liquid crystal at the substrate-droplet interface. Cracks and allied surface instabilities are detrimental to the quality of the final product like surface coatings, and therefore, its suppression by an external inert additive is a promising technique that will be of immense importance for several industrial applications. We believe this fundamental investigation of crack suppression will open up an entire avenue of applications for the NLCs in the field of coatings, broadening its already existing wide range of benefits.

Evaporation induced self-assembly of colloidal nanoparticles from droplets of colloidal suspensions hold extensive significance in the surface coating and ink-jet printing industries. However, formation of cracks in these dried colloidal films [1] compromises the product quality, and therefore, significant efforts have been directed towards developing techniques that lead to fabrication of crack-free [2–4] or crack-suppressed [5,6]colloidal films. These techniques can be broadly divided based on the mechanism that target a particular component of the colloidal system (i.e. the colloidal particles [7], the substrate [8,9] or the solvent [10,11]) besides additional techniques such as alterations in the drying methodologies [12,13] and/or externally induced perturbations [14] that are inherent to the drying process itself. Moreover, a set of alternative mechanisms exist that involves introduction of a foreign component to this parental colloidal system to forcefully suppress crack formation like addition of sol-gel [2] or non-adsorbing polymers, soft spheres [10]etc. It is, however, cumbersome to control the substrate's physical properties like elasticity of a polydimethylsiloxane (PDMS) coating that involves mixing of base and cross-linker in pre-determined ratio. A major drawback of the aforementioned techniques reported previously lies in their inability to *recover the additive from the crack-free colloidal film*.

Here we propose a unique additive – nematic liquid crystal (NLC) droplet that eliminates crack formation and produces crack-free films that can be easily separated from its additive (i.e NLC) for further processing. Moreover, we have investigated the applicability of this additive across a wide range of substrates in terms of its varied surface and mechanical properties. We further attempt to provide the fundamental mechanism that gives rise to complete suppression of cracks on introduction of NLC droplets with respect to the major forces governing the phenomenon.

For the present study, colloidal droplets (2μl, 1(w/w)%, particle diameter: 48nm, surface modified with carboxylate groups) are placed on substrates of varying surface properties (Please note, substrate preparation and characterization of the fabricated substrates are provided in the **Supplementary Information**, Note S1, Table S1) and are subjected to natural evaporation. The entire droplet evaporation process is observed using an optical microscope and it is found that the loss of solvent from the droplets gradually transforms them into thin colloidal films. These films succumb to the capillary pressure induced by the evaporation and undergo disintegration to form prominent cracks. This process of crack formation over substrates of varying wettability [16] and specific elasticity [9] are characterized extensively. Therefore, we consider crack formation on a substrate of specific surface property (wettability) and mechanical property (elasticity) as the control/reference for the present investigation.

The experimental procedure comprises of determination of the quantity (by volume) of NLC additive to be added to eliminate the possibility of crack formation. For this purpose, we have used the NLC 5CB (4-Cyano-4'-pentylbiphenyl) (procured from *Sigma Aldrich*). Thus, sessile droplets containing dispersed particles (as mentioned previously) of identical volume and particle concentration are placed on a hydrophilic rigid substrate. To these droplets, varying volumes (0.1µl, 0.2µl, 0.3µl, 0.5µl) of NLC suspension are added and this colloid-NLC droplet is kept undisturbed till cracks are visible on the dried colloidal film.

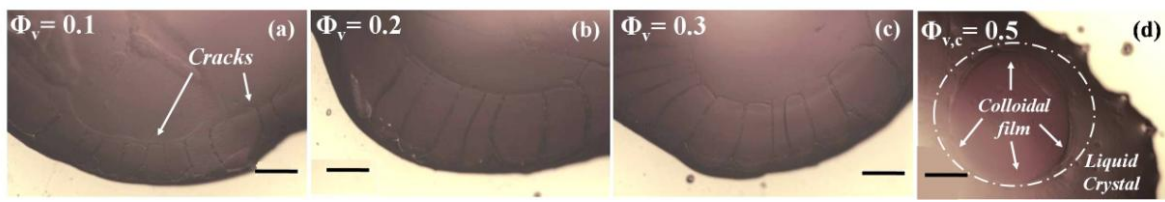

**Figure 1:** Optical microscopy images **(a)-(b)-(c)** show the droplet behaviour with an increasing volume of liquid crystal ($\Phi_v$ in µL). It clearly indicates crack formation at the dried film edge till a critical volume of LC is added. **(d)** Represents the primary suppression of crack formation pertaining to a critical volume ($\Phi_{v,c}$) of LC. The dried colloidal film in **(d)** is devoid of cracks and is surrounded by a pool of liquid crystal suspension. **Scale Bar**: 200µm.

**Figure 1** shows a sequence of optical microscopy images that outline the effect of volume of NLC on crack suppression. **Figure 1(a)-(b)-(c)** show that the cracks exist at the edge of dried colloidal film even on increasing the LC volume from 0.1 µL to 0.3 µL whereas in the decisive image presented in **Figure 1 (d)** these cracks are absent. It is evident that the NLC can flawlessly eliminate cracking although its crack suppression ability is effective only beyond a critical volume. For the present experimental parameters and materials (substrate-particle combination), this critical volume ($\Phi_{v,c}$) of NLC is ~ 0.5µL. Hereafter, the experiments have been performed at the critical volume of NLC and the results presented hereafter (if not mentioned specifically) correspond to the critical volume of NLC additive.

***Recovery of NLC*** - The weight of several colloidal droplets of identical volume (~2 µL) were measured using a high precision mechanical balance (*Sartorius*, Gottingen, Germany.) at the start of the evaporation (drying experiment) and was found to be ~0.0027mg. Similarly, the weight of the added NLC ~ (~0.5µL) is measured and is found to be ~0.0010mg. Post drying, the NLC layer is extracted with extreme care using a micro-pipette and the colloidal film settles on the substrate. Weight of these dried colloidal films is found to be close to zero implying the solvent has almost evaporated. This provides immense superiority to the present technique in terms of ease of additive recovery, re-usability of the additive and production of a crack-free colloidal film at in absence of external heating. The technique is eco-friendly and can be touted as the '*evaporative lithography*' of the future to produce crack-free colloidal films (Please refer to **Supplementary Information, Note S2** for a representative scanning electron micorscopy image of the recovered dried colloidal film)

*Effect of Substrate Wettability* – To outline the effect of substrate wettability on the crack suppression strength of NLC, evaporation experiments are performed on substrates of specific wettability – hydrophilic and hydrophobic. The frequency of cracks and morphology of the final crack pattern are known to be heavily influenced by the substrate wettability [16]. This is clearly seen in the typical trench, minimal (~2) cracks on the hydrophobic substrates (**Figure 2(a1)**) which, in contrast, are quite dissimilar to the arch shaped, multiple cracks ((**Figure 2(a2)**) appearing on the hydrophilic substrates, both in absence of any additive.

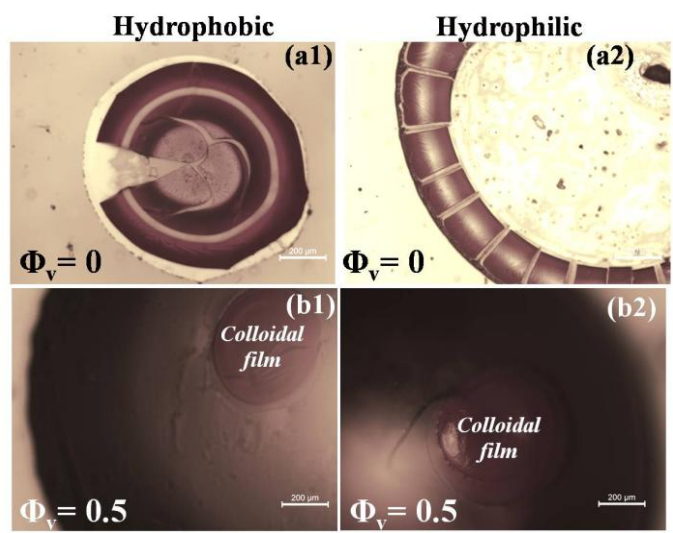

**Figure 2:** Characteristic crack morphology on substrates of distinct wettability **(a1)** hydrophobic and **(a2)** hydrophilic. Formation of crack-free colloidal films on addition of NLC on **(b1)** hydrophobic and **(b2)** hydrophilic substrates.

Unlike the nature of the colloidal films seen in **figure 1**, addition of the NLC additive to the evaporating colloidal droplets on hydrophobic and hydrophilic substrates lead to the formation thin colloidal films that are completely crack-free (**Figure 2(b1,b2)**). Formation of crack-free NLC-colloidal film droplets are found to be highly repeatable across surface wettability. Thereby, NLC addition as a technique to obtain crack-free films can be applied irrespective of the wetting state of the substrate. This substrate wettability independent crack suppression property of NLC is unique and increases its applicability beyond traditional domains. For obtaining further insights into crack suppression process by NLC drop, we proceed to delve into real-time monitoring of the evaporation phenomenon on a hydrophobic substrate using in-situ confocal microscopy.

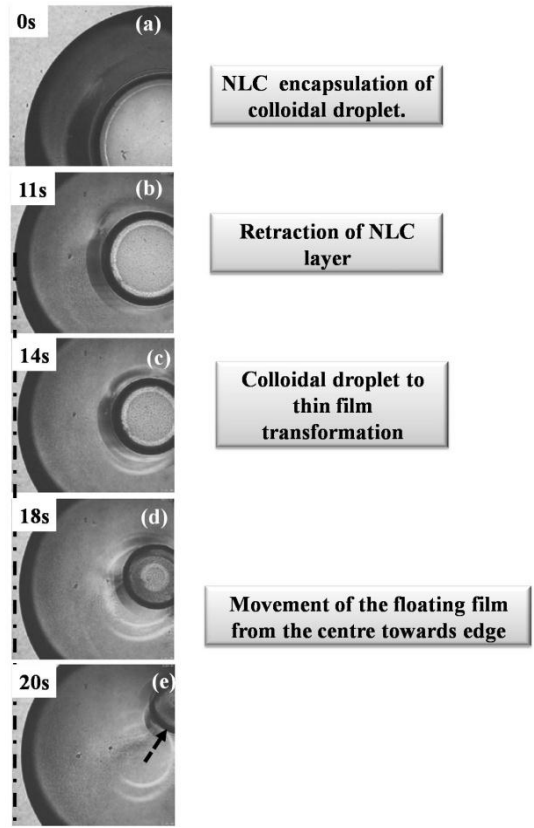

**Figure 3:** Dynamic evolution of the crack suppression on addition of LC to a colloidal droplet on a hydrophobic substrate using in-situ confocal microscopy.

Here we observe that in the initial stage of evaporation, NLC forms a thin protective layer over the colloidal droplet. Formation of this thin layer is instantaneous and can be easily demarcated since NLC has a characteristic gray colour whereas the colloidal droplet is differently coloured (pink). The colloid-NLC interface is distinct as marked in **Figure 3(a).** It must be mentioned here that the refractive indices of the colloidal droplet and NLC layer differs considerably and this is a major obstacle in accurate determination of the thickness of the NLC envelope. This primal stage of encapsulation of the colloidal droplet by the NLC layer is followed by the gradual retraction of (**Figure 3(b),3(c)**) the NLC layer. This NLC layer then slowly forms a cushion for the evaporating colloidal droplet that simultaneously shrinks to form a thin colloidal film. The 'relatively lighter' colloidal thin film begins to floats over the NLC cushion. The last stage is the most intriguing part where in the floating colloidal film at a certain instant veers from the centre of the NLC cushion ((**Figure 3(d), 3(e)**) towards the edge. Surprisingly, the film attains an equilibrium position that is neither at the centre of the LC cushion nor at it's the edge; instead, it floats somewhere between these extremes (Real time movie of the evaporating colloidal droplet dynamics is available in the **Supplementary Information**, Movie S1). A pertinent question at this juncture that may arise is whether the floating condition is a mere manifestation of the density difference between the NLC and the colloidal droplet (buoyancy effect). However, the density of polystyrene particles and NLC are comparable ~1.02 kg/m$^3$ and nullifying the buoyancy effect at the onset. We performed an additional control experiment wherein the colloidal droplets (~2 µL) are allowed to evaporate with heavier (denser) liquid (FC-40 oil) of equivalent amount of that of NLC (~ 0.5 µL) on a hydrophilic substrate and the results are presented in **Figure 4**. It can be seen that cracks appear in case of denser fluid (FC-40), and therefore, the action of NLC is definitely beyond the buoyancy effect.

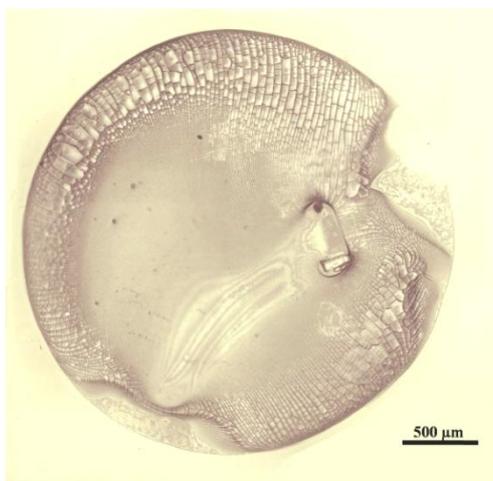

**Figure 4** Effect of fluid density on the crack suppression ability of NLC. Dried colloidal films with addition of FC-40 of equivalent amount ~0.5 μL is filled with cracks.

Now, we attempt to qualitatively describe the situation observed above. Two noteworthy results must be mentioned towards the effort in understanding the underlying mechanism of the present phenomenon: firstly, the colloidal drop undergoes rapid evaporation after it is placed on the glass substrate; and secondly, the NLC drop having a large vapour pressure seems not to evaporate – it remained as a perfect drop even after a week of placing it on the glass substrate. As a consequence, the colloidal drop realizes a reduced volume being subjected to strong evaporation, and thereby, floats at the surface of the cushioning NLC drop. Below a certain radius ratio in a compound drop situation, the drop having less volume floats midway over the sessile (here NLC) drop [17]. However, LCs tend to show a high energy defect spot at the hemispherical tip [18,19], thereby, facilitating an off-centre shift of the colloidal drop as has been observed above. Moreover, such a quick drift of the colloidal drop may not occur if the defect core remains absent – a scenario which may arise depending on the rigidity of the substrate on which the NLC sessile drop rests. Towards this, we proceed to investigate the aforementioned mechanism on soft elastic substrate.

*Effect of Substrate Elasticity-* Finally, we focus our attention to investigate the above mechanism on substrates with different elasticity ranging from a rigid substrate (that is the erstwhile PDMS coated hydrophobic substrate) to a relatively softer substrate fabricated by altering the base: cross-linker ratio (30:1) of PDMS [20]. These substrates have been characterized in terms of their Young's modulli of elasticity and coating thickness (Details in the **Supplementary Information**, Table S1). Thereafter, fresh evaporation experiments are performed and the rigid substrate utilized for evaporation of pure colloidal droplets is the reference for outlining the effect of substrate elasticity on the crack suppressing capacity of NLC by comparing them with the results obtained on softer substrates.

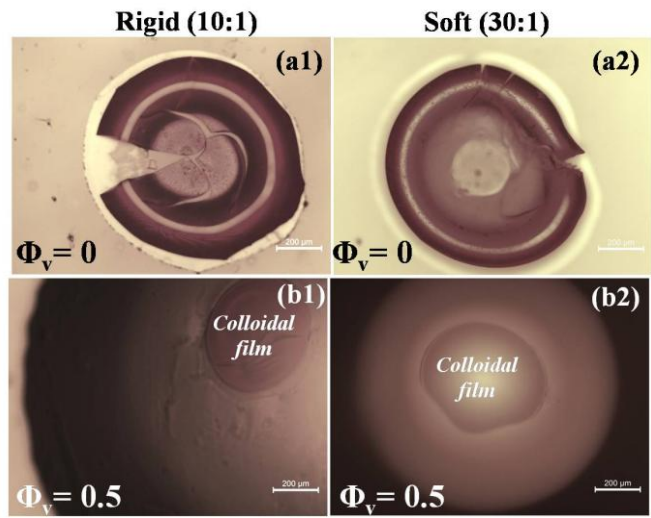

**Figure 5:** Final morphology of the colloidal film in absence of NLC on a **(a1)** rigid substrate **(a2)** soft substrate, respectively. The equilibrium location of the colloidal film is off-centre **(b1)** on the rigid substrate and at the centre **(b2)** of the underlying NLC layer over a relatively softer substrate.

It is evident from **Figure 5** that NLC exhibit surface-elasticity independent crack suppressing ability which can be employed to obtain a crack-free colloidal film even on softer substrates.

A careful observation of the Figures **5(b1)** and **(b2)** brings out the striking difference in terms of the location of the floating colloidal film on the NLC cushion. This implies that the stages of crack suppression in case of softer substrates are altered. We resorted to in-situ confocal microscopy to study the dynamic evolution of the colloidal droplet on addition of NLC over a soft substrate. Interestingly, the stages observed over a rigid hydrophobic substrate that is formation of a NLC protective layer, gradual retraction of this envelope, detachment of the colloidal droplet from the substrate, and subsequently, afloat condition of the colloidal film over the NLC cushion are exactly identical over softer substrates (Real time movie of the evaporating colloidal droplet dynamics in the **Supplementary Information**, Movie S2). However, the equilibrium position of the colloidal film is at the centre of the NLC cushion and continues to remain at the centre (even after 24hours) for the entire process as well as thereafter. Thus, the stage of swift movement of the colloidal film from the centre of the NLC cushion to its edge is invariably absent on soft substrates. As explained earlier, such an observation may be attributed to the defect-free hemispherical core that exists on NLC sessile drop placed on soft elastic substrates since it can deform the soft surface to minimise its elastic energy. *In other words, the equilibrium position of the colloidal thin film can be tuned based on the substrate rigidity when drying occurs under the influence on NLC.* This insinuates towards yet another important facet of the process, that the final location of the colloidal film in an NLC-colloid binary system is tailored or governed by the arrangement of the NLC at the colloid-NLC interface.